\documentclass[preprint,12pt]{elsarticle}

\usepackage[]{algorithm2e}
\usepackage{amssymb}

\journal{Radiation Physics and Chemistry}

\begin{document}

\begin{frontmatter}



\title{A Geometric Approach to Spectral Analysis}


\author[1]{Silvio B. Melo}
\author[2]{Ilker Meri\c c}
\author[3]{Fabiano B. M. Silva}
\author[1]{Carlos C. Dantas}
\author[2]{Jarle R. S\o lie}
\author[2]{Geir A. Johansen}
\author[4]{Bj\o rn T. Hjertaker}
\author[3]{Bruno J. S. Barros}

\address[1]{Universidade Federal de Pernambuco -Recife-Brazil }

\address[2]{ Western Norway University of Applied Sciences – Bergen - Norway}
                    
\address[3]{Universidade Federal Rural de Pernambuco Garanhuns-Brazil}

\address[4]{University of Bergen - Bergen - Norway}

\begin{abstract}
Analyses of gamma-ray spectra, acquired through non-invasive techniques, have found applications in fields such as medicine, industry and homeland security. Constituent gamma-ray spectra of a chemical compound have been determined from its sole spectrum through a forward Monte Carlo simulation coupled with a least squares method (MCLLS).  The method's limitations include its linearity assumption and its oversensitivity to correlated or noisy data, which render the method unfit to deal with such numerical ill conditioning. Recently this issue was tackled by iteratively reducing the condition number of the linear system of equations. Despite its superior results, it is not suitable for cases where there are missing libraries in the analysis. Our work introduces a novel framework that allows treating spectral analyses problems through geometrical insights. Based on this it was possible to propose solutions to three problems regarding the missing library: to find its photopeak, its most probable fraction, and an envelope around its spectrum. We successfully validated these on some Monte Carlo-generated radionuclide gamma-ray spectra.

\end{abstract}


\end{frontmatter}


\section{Introduction}
\label{}
The unfolding of gamma-ray spectra is a commonly encountered aspect of inverse radiation analyzers that have found applications within a variety of fields such as medicine, industry as well as homeland security (Ali and Rogers, 2012\cite{Ali}; Borsaru et. al. 2006\cite{Borsaru}; Im and Song, 2009\cite{Im}; Meric et. al., 2011\cite{Meric11}; Molnàr, 2004\cite{Molnàr}; Zhang and Gardner, 2004\cite{Zhang}). A good application example where the unfolding of gamma-ray spectra is required is within the inverse analysis of bulk samples carried out using the well-known technique of prompt gamma-ray neutron activation analysis (PGNAA) (Gardner et. al., 1997\cite{Gardner97}). The PGNAA is a widely used elemental analysis technique that has found numerous applications within the industry as it provides a means of on-line, rapid and non-intrusive interrogation of bulk samples. The compound nuclei formed upon thermal neutron capture reactions in the sample will almost instantaneously decay into a more stable state through the emission of highly energetic gamma-rays. These gamma-rays are characteristic of every element and in fact, of every isotope of every element allowing detailed elemental analysis of bulk samples. 

In conventional applications of PGNAA, the so-called single peak analysis technique is utilized for the quantitative analysis. The single peak analysis technique ultimately relies on finding the most intense photopeaks in the measured prompt gamma-ray spectra, identifying the constituents of a bulk sample based on the energies of these gamma-rays and using the overall intensities to finally determine the elemental amounts (Gardner et. al., 1997\cite{Gardner97}). This is still a common approach that does not require any form of spectrum unfolding, however, it indeed requires the use of gamma-ray detectors with extremely energy resolution and preferably also good detection efficiencies. Most importantly, the technique disregards much of the spectral information contained within the Compton continua of each pertinent photopeak. To eliminate the above mentioned drawbacks of the single peak analysis technique, the so-called Monte Carlo Library Least-Squares (MCLLS) technique has been proposed and applied successfully in many applications of the inverse PGNAA analysis of bulk samples (Gardner et. al. 2005\cite{Gardner05}; Meric et. al. 2011\cite{Meric11}; Wang et. al., 2008\cite{Wang}). Here, it should be noted that a “library” refers to the prompt gamma-ray spectrum of a single constituent. Briefly, the MCLLS approach can be summarized to be consisting of the following steps:

\begin{enumerate}
\item Obtain or assume an initial composition of the sample being investigated
\item Obtain individual elemental libraries, i.e. spectra, through accurate forward Monte Carlo (MC) simulations
\item Execute a library least-squares (LLS) search to obtain the library multipliers which are then used to calculate the amounts of each constituent in the sample
\item If the calculated amounts are so far apart from the assumed sample composition, then go back to step 2 and repeat steps 2 and 3 using the calculated amounts as the new sample composition in the next iteration.
\end{enumerate}

The MCLLS approach has several advantages over the single peak analysis; unresolved peaks are treated automatically and the entire spectra, including their Compton continua are used in the subsequent quantitative analysis. However, it has also been recently shown in a previous work (Meric et. al. 2012) that the MCLLS approach will suffer from “ill-conditioning” which may be caused by a number of factors such as:

\begin{itemize}
\item two or more libraries used in the quantitative analysis are similar in shape, i.e. when two or more libraries are linearly correlated,
\item a negligible contribution from a certain library to the overall number of counts in the total prompt gamma-ray spectrum which may be due to presence of trace elements or elements with extremely low neutron capture cross-sections in the bulk sample.
\end{itemize}

A detailed account of the ill-conditioning in the MCLLS approach used in conjunction with inverse radiation analyzers and a proposed solution to this problem is given elsewhere (Meric et. al., 2012\cite{Meric12}). It should be mentioned, however, that the proposed treatment consisted of an iterative method based on constructing combinations of libraries that would minimize the condition number of the linear system of equations formed in the MCLLS approach. The proposed method has successfully been applied to the highly ill-conditioned case of the multiphase flow measurement where the sample consisted of homogeneous mixtures of oil, gas, water and salt (Meric et. al., 2014\cite{Meric14}). This can be considered to be a particularly difficult case for the MCLLS solver as both water, oil and gas phases contain hydrogen and also due to the fact that oil and gas phases are simply made up of the same elements, i.e. hydrogen and carbon, and will thus have the same spectral shapes.

The proposed iterative method does indeed provide a means of minimizing the ill-conditioning in the MCLLS approach utilized for the quantitative analysis. The method does, however, assume that all constituents of a given sample are identified and that the libraries for each of the constituents are generated and available prior to the analysis. This is not always the case, especially in practical applications where missing libraries may not be avoided due to incomplete knowledge about the sample itself, detector activation as well as incomplete knowledge about the material composition in the surroundings of a given experimental setup.

Therefore, the present work focuses on identifying missing libraries in the MCLLS approach. For this purpose, a novel barycentric geometrization of the linear problem is proposed and its feasibility is explored through using somewhat simpler MC generated radioisotopic gamma-ray spectra. For the generation of these spectra, the general purpose MC code system MCNP6 (Goorley et. al., 2012\cite{Goorley}). It should be emphasized that, although the main motivation is due to the MCLLS approach, the method is sufficiently general to be applied in other applications of spectrum unfolding.  

In the remainder of the manuscript, an introduction to the above mentioned geometric modeling will be given. This will be followed by a description and presentation of the results of the initial numerical experiments carried out to prove the feasibility of the geometric modeling. Finally, some conclusions will be drawn based on the results of the pertinent numerical experiments.

\section{Geometric Modeling}

\begin{figure}
{\center
\includegraphics[scale=0.40]{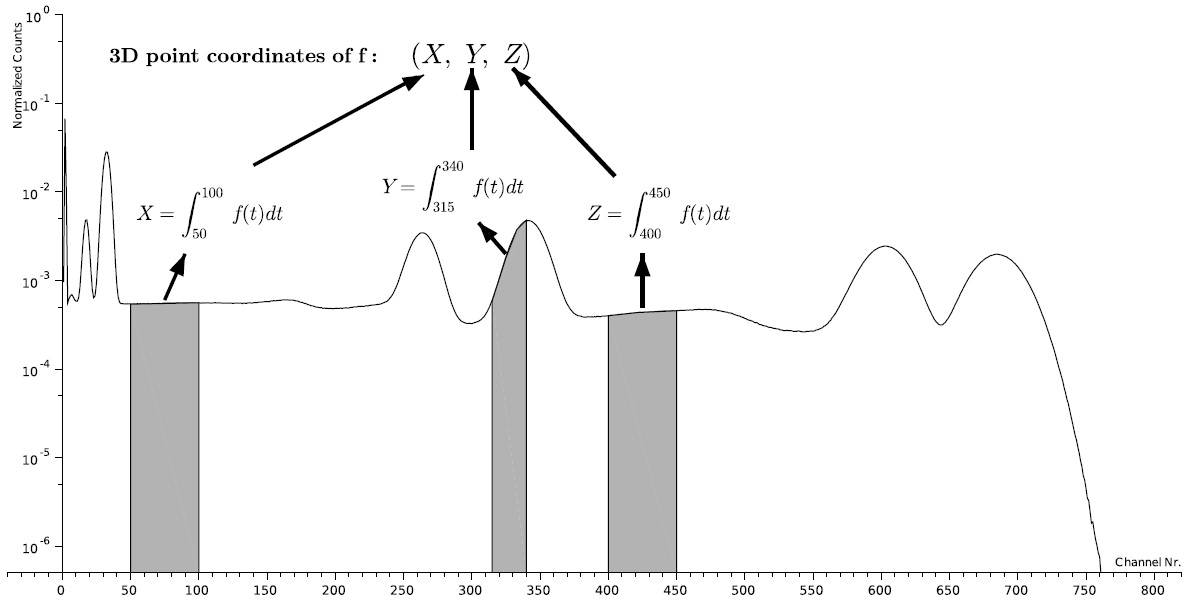}
\caption{A library spectrum is represented as a 3D vector by integrating the counts along three fixed intervals. The intervals choice must be the same for all libraries.}\label{3Dvec}
}
\end{figure}

The existing operational relationship between geometry and algebra has brought immense benefits to both realms. Modeling algebraic problems in such a way as to allow geometric tools to be applied establishes a very effective visual framework to attack these problems, although in some cases it may introduce unwanted biases. Linear treatment of spectral analysis is amenable to a geometric modeling considering that: 
\begin{itemize}
\item the elemental spectrum is a discretization of a continuous function of a given interval; 
\item the set of spectra spanned by basic compounds is a vector space; and 
\item the amount of basic compounds is finite.
\end{itemize}
 The aforementioned discretization involves partitioning an interval in equally spaced subintervals, and taking a sample from each subinterval (channel), which is associated to an energy level by the spectrometer, usually NaI or BGO scintillation detectors (Im and Song 2009\cite{Im}). The set of basic elements being small and the unknown substance being generated as a linear combination of the basic compounds mean that only a small amount of appropriately chosen channels needs to be considered. If there were no noise or missing element the amount of appropriately chosen channels would correspond to the amount of basic elements. What is meant by “appropriately chosen” channels is that they should be picked in such a way as to ensure linear independence to the spectra of the set of basic elements. The main goal of this process is to extract from the set of spectra the minimum information needed to determine the mass fractions of the constituents of a certain compound, assuming that the spectra of these constituents are available. However, other problems around this issue do exist: how to minimize noise influence, how to estimate and avoid anomalous photon counts (from scattered radiation), how to minimize the influence of ill-conditioning, how to minimize the influence of non-linearity, how to isolate missing constituents, etc. In this work a geometric modeling is proposed that amounts to a framework with which one can tackle different types of problems with the help of visualization and the structural coherence with the geometry. The visualization is possible for cases with three constituents, but even with higher amounts we can use geometric configurations from the 3D case that can be generalized to higher dimensions.
We start by considering that a spectrum sampled at n energy levels (channels) can be seen as a point of $I\hspace{-1mm}R^n$, but because the amount of basic constituents is small, there is no loss of information by considering just the (affine) space spanned by these basic constituents through linear combinations, which is isomorphic to $I\hspace{-1mm}R^k$, where $k$ is this amount. The unknown compound is just a point in this space. For the case where there is a possibility of a missing element in the prescribed set of constituents, we can consider one extra dimension to be able to detect this situation. If, for instance, three basic constituents are to be considered and noise is not an issue, one can find three channels in such a way that the basic elements’ values at these channels would make them a basis for the three-dimensional space. Figure \ref{3Dvec} shows how to produce a vector out of a spectrum, given certain interval choices.

\begin{figure}
{\center
\includegraphics[scale=0.60]{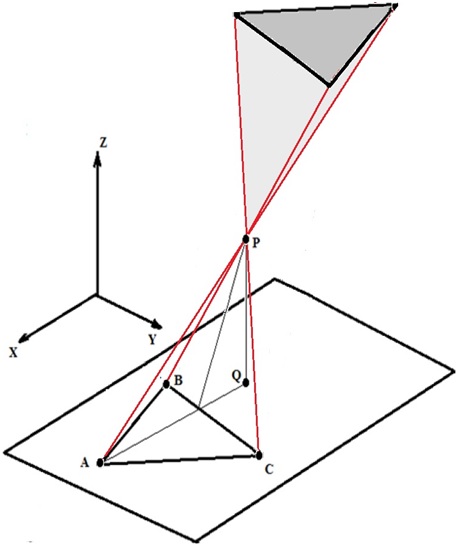}
\caption{Affine plane spanned by points A,B and C. Point P is outside this space, and Q is the plane's best approximation to P. The grey region is where lies a point that, together with A, B and C, yield a tetrahedron containing P. It is where any missing library's point representation must lie.}\label{AffPlane}
}
\end{figure}

From three constituents we produce three vectors, all based upon exact same interval choices. Since what is required to solve in this problem is to find the weight fraction of each basic element with respect to the unknown compound, only $barycentric\ combinations$, among all liner combinations, should be used to produce new compounds interpreted in this way. A barycentric combination is one in which the coefficients must add to one, similar to a weighted average. This restriction imposed on three constituents corresponds to a point lying on an affine plane of $I\hspace{-1mm}R^3$ (the ``affine'' term is because the plane does not necessarily includes the origin). In this case a three-vector can be interpreted as a point in three-space, and the affine plane spanned by these three points can be interpreted as containing the representations of all possible elements (fictitious and non-fictitious) that can be produced from combining the basic components in all possible proportions. Points that are inside the triangle are obtained through the use of non-negative coefficients in the barycentric combination, and they correspond to all physical (non-fictitious) elements that can be produced from the three basic elements. In Figure \ref{AffPlane}, points $A$, $B$ and $C$ depict valid constituents, which are inside a quadrilateral that portraits an affine plane of $I\hspace{-1mm}R^3$.

In general, if we have $k$ basic elements, then we fix $k$ intervals of channels, i.e., they are elements of $I\hspace{-1mm}R^k$, then we produce the corresponding $k$-points by integrating their spectra over the intervals. The affine space spanned by these points is $(k-1)$-dimensional, and a point $Q$ lying on this affine space can be written as: 

\begin{equation}\label{Baryc}
Q=\sum_{i=1}^{k} \alpha_i\cdot P_i\ 
\end{equation}
 
where $\sum_{i=1}^{k} \alpha_i=1$ and $P_i$ are the $k$-points built from the spectra. In this case the produced physical elements corresponding points are located inside the $(k-1)$-dimensional simplex having  $P_i$ as vertices. The point $Q$ is outside this simplex if and only if there is an index $j$ such that $\alpha_j<0$. Figure \ref{AffPlane} illustrates this situation for $k=3$. If, however, there is a point which is outside the $(k-1)$-dimensional affine space spanned by the points, then Equation \ref{Baryc} becomes unsolvable. This is the case of point $P$ in Figure \ref{AffPlane}. Point $Q$ is the closest one to $P$ which is still inside the affine space spanned by $A, B$ and $C$, i.e., point $Q$ is the least squares solution to Equation \ref{Baryc}. Outliers like $P$ may appear when the data is too noisy, or when at least one of its constituents is neglected.  Let us consider for a moment that $P$ is a point corresponding to a compound for which there is a constituent, say $D$, whose spectrum is unknown and, therefore, it was not listed among the basic constituents. Then we can write $P$ as a barycentric combination of all four constituents:  $P=\alpha\cdot A+\beta\cdot B+\gamma\cdot C+\delta\cdot D$, where $\alpha+\beta+\gamma+\delta=1$. The four points now can span the entire $I\hspace{-1mm}R^3$ through such combinations, but only points confined in their 3-simplex (tetrahedron) are physically plausible, with positive coefficients. This observation gives us a way of limiting the region where $D$ should lie, by connecting each of the basic points $A, B$ and $C$ to $P$, creating a triangular pyramid (grey region) above $P$ in Figure \ref{AffPlane}. If we assume that the margin for the presence of missing constituents is relatively narrow, then we can limit this region even more (ex: missing libraries should be no more than 10\% of the total compound). This feature can be generalized to higher dimensions.

\section{Method and Experiments}
In the present work, we use the above developed geometric modeling to infer useful information on the missing constituent. The most fundamental question to start off this method is how to determine the most appropriate intervals of channels.  Simulations performed with three and four constituents, with interval determination conducted via a Tikhonov regularization (Silva et. al 2016\cite{Silva}) have shown that the best possible choices must correspond to the photopeak regions of each spectrum.  In this case, each interval was modeled as a sliding window and, for each configuration of intervals so defined, the least squares solution of Eq. \ref{Baryc} was computed. Since this is subject to numerical imprecisions and noise, this solution was regularized by imposing a reduced conditioning on the matrix of the system. It was also observed the graphs of the linear correlation coefficient between each pair of elements and the smallest singular value of the matrix, they all indicating that the highest linear independence (the closest to orthogonality) occurred around photopeaks regions. Since the initial constituents are given, their photopeaks regions are known in advance. 

\subsection{Implementation}
With the above stated as a starting point, we can enunciate the three steps of our proposal: 
\begin{enumerate}
\item Find the missing constituent's photopeak; 
\item Find the most probable proportion of the missing constituent; 
\item Build an envelope around the missing constituent’s spectrum.
\end{enumerate}
For the first step, we increment the dimension of each point by one, and the added dimension corresponds to an interval whose positioning is to be found. The idea is to solve Eq. \ref{Baryc} as the added interval slides along the energy axis. The search is conducted as to maximize the residue of the least squares solution, since we need to find the region where the least squares solution becomes the most distant from the compound point representation, which happens supposedly when the photopeak of a missing constituent is found (see Algorithm 1). In the Algorithm 1, $V_P$ is the vector of the total compound, $V_{L_j}$ is the $j^{th}$-library' $k$-vector and $I_{size}$ is the size of the interval window.

In the second step, we build a new constituent to act as the missing library, by assigning count values for it at all intervals, including the added one. This time, for each assigned value at each interval, a least squares problem for Eq. \ref{Baryc} (with the added constituent) is solved, where the residue is to be minimized. The solution of the least squares associated with the smallest residue contains the mass fractions of all constituents, particularly the fraction of the new constituent, which should be an approximation of the fraction of the missing library (see Algorithm 2). In the Algorithm 2 the missing library is referred to as $L_k$.

In the third step, we use the idea presented in the previous section, which produces straight lines that connect each basic constituent's point to the compound's corresponding point. Suppose the correct fraction of the missing library is $\delta$. Since in an edge all barycentric coefficients are zero, with the exception of the simplex points that lie in this edge, the barycentric expression of this edge becomes: $P=(1-\delta)\cdot P_i+\delta\cdot U$, where $P$ is the compound, $P_i$ is a basic constituent and $U$ is the missing library. Then we can write $U=(1/\delta)\cdot P-((1-\delta)/\delta)\cdot P_i$, defining a vertex in the truncated pyramid where the missing library should lie.  This is done for each basic constituent, and for a range of $\delta$: $[\delta-\epsilon, \delta+\epsilon]$, where $\epsilon$ is a tolerance value, that ultimately determines the envelope's width (see Algorithm 3). In the Algorithm 3, the subscript $c$ stands for the current channel's index of any given library's spectrum.	\\

\begin{figure}
{\center
\includegraphics[scale=0.42]{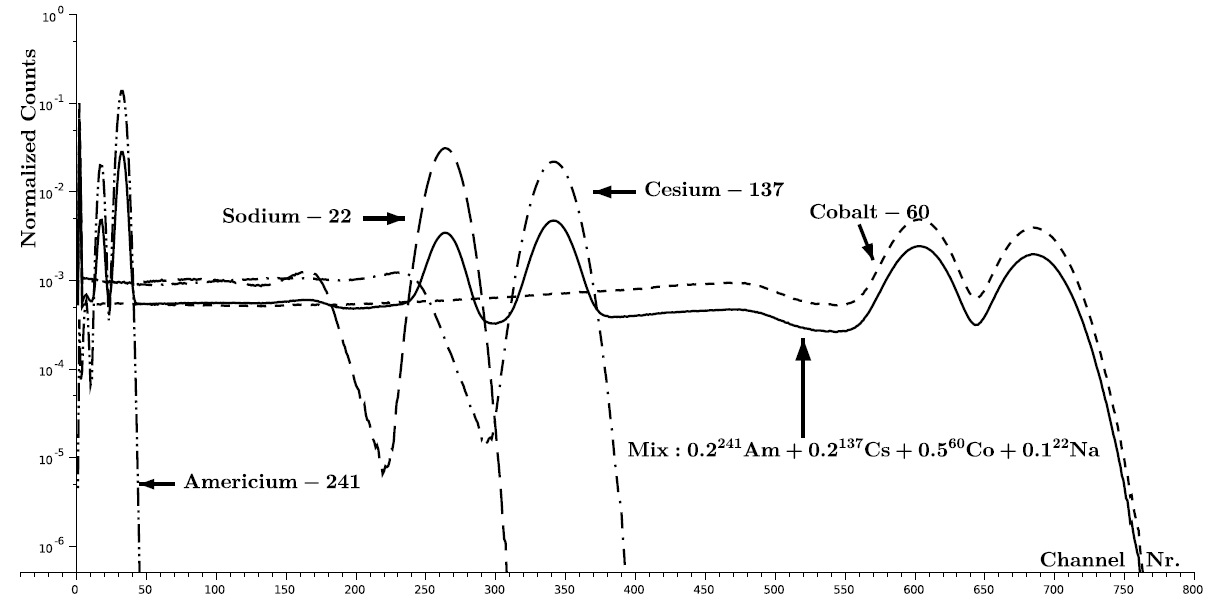}
\caption{Five examples of Monte Carlo generated gamma-ray library spectra, together with a summed spectrum – 20\% Am, 20\%Cs, 50\%Co and 10\%Na.}\label{Gammaspectra}
}
\end{figure}

\begin{algorithm}[]
\SetAlgoLined
\KwResult{maxRESchannel is the photopeak interval's lower limit }
{\bf begin}\\
\ \\
 increment vectors dimension from $k-1$ to $k$\;
 initialize $k^{th}$ dimension's interval $I_k$ with $[1,I_{size}]$\;
 
\ForEach{library's $k$-vector $V_{L_j}$}{
   initialize the $i^{th}$-coordinate of $V_{L_j}$ by integrating $L_j$'s spectrum over interval $I_i$, with $i\in\{1,...,k-1\}$\;
 }
 maxRES$\leftarrow$ 0\;

\ForEach{channel $c$}{
   $I_k \leftarrow [c,c+Isize]$\;
   update the $k^{th}$-coordinate of all libraries $k$-vectors by integrating their spectra over the interval $I_k$\;
   solve $V_P=\sum_{j=1}^{k-1} \alpha_j\cdot V_{L_j}$ with Least Squares and put its residue in LSresidue\;
   {\bf if} maxRES $<$ LSresidue {\bf then}\\
   \ \ \ maxRES$\leftarrow$LSresidue\;
   \ \ \ maxRESchannel$\leftarrow c$;
 }
{\bf end}\\
 \caption{Finding the missing library's photopeak.}
\end{algorithm}

\begin{algorithm}[]
\SetAlgoLined
\KwResult{The estimated missing library's $k$-vector $M$}
{\bf begin}\\
\ \\
 
\ForEach{library $L_j$, $j\in\{1,...,k-1\}$}{
   initialize the $i^{th}$-coordinate of $V_{L_j}$ by integrating $L_j$'s spectrum over interval $I_i$, with $i\in\{1,...,k\}$\;
 }
 minRES$\leftarrow$ $\infty$\;

\ForEach{$V\in [minCount,maxCount]^k$}{
   $V_{L_k}\leftarrow V$\;
   solve $V_P=\sum_{j=1}^{k} \alpha_j\cdot V_{L_j}$ with Least Squares and put its residue in LSresidue\;
   {\bf if} minRES $>$ LSresidue {\bf then begin}\\
   \ \ \ minRES$\leftarrow$LSresidue\;
   \ \ \ $M\leftarrow V$;\\
   {\bf end}
 }
{\bf end}
 \caption{Estimating the missing library's point representation.}
\end{algorithm}

\begin{algorithm}[H]
\SetAlgoLined
\KwResult{The envelope around the missing library's spectrum}
{\bf begin}\\
\ \\
 initialize each channel of $minEnv$, the lower envelope, with $\infty$\;
 initialize each channel of $maxEnv$, the upper envelope, with $0$\;
 $maxFrac\leftarrow$ missing library's fraction $+\epsilon$\;
 $minFrac\leftarrow$ missing library's fraction $-\epsilon$\;
\ForEach{basic library $L_j$,with $j\in\{1,...,k-1\}$}{
   \ForEach{channel $c$}{
   $\delta\leftarrow minFrac$\;
    $U_c\leftarrow(1/\delta)\cdot P_c-((1-\delta)/\delta)\cdot L_{j_c}$\;
    {\bf if} $minEnv_c > U_c$ {\bf then}\\
     \ \ \ $minEnv_c\leftarrow  U_c$\;
   $\delta\leftarrow maxFrac$\;
    $U_c\leftarrow(1/\delta)\cdot P_c-((1-\delta)/\delta)\cdot L_{j_c}$\;
    {\bf if} $maxEnv_c < U_c$ {\bf then}\\
     \ \ \ $maxEnv_c\leftarrow  U_c$\;
 }
 }

{\bf end}\\
 \caption{Estimating the missing library's envelope.}
\end{algorithm}

\subsection{Results}
The method was tested with five Monte Carlo-generated gamma-ray library spectra: Americium-241, Sodium-22, Cesium-137 and Cobalt-60 (see Figure \ref{Gammaspectra}). The gamma-ray spectra were generated using point like sources for a 3"x3" Nai detector using MCNP6. For this purpose, the so-called pulse-height tally of MCNP6 was used in conjunction with the Gaussian Energy Broadening (GEB) option to produce relatively more realistic gamma-ray spectra. In each simulation a total of $10^8$ primary photons were simulated, thus the statistical uncertainty in each simulation was negligibly small. 
 To test the first step, we first regarded $^{22}$Na as a missing library, with a fraction of 10\% in the total, and the rest as basic constituents, with proportions: 20\% of  $^{241}$Am, 20\% of $^{137}$Cs, and 50\% of $^{60}$Co. Their photopeaks intervals used were: 10 to 45 for $^{241}$Am, 293 to 395 for $^{137}$Cs, and from 544 to 764 for $^{60}$Co. We assumed only that the missing library's interval width was 60. The Algorithm 1 was tested with this data and returned the interval $[235,295]$, which corresponds to the actual photopeak of $^{22}$Na. Then we tested it with $^{241}$Am as the missing library, with the result $[5,65]$, and afterwards we tested $^{137}$Cs as the missing library, resulting in interval $[311,371]$, which are the best solutions for their photopeaks with interval size 60. Next, we took $^{22}$Na again as the missing library, to test the robustness of the method, starting with its regular proportion of 10\% and reducing it, each time one order of magnitude smaller, while compensating this reduction by increasing the proportions of the others uniformly.  The method was capable of finding $^{22}$Na's photopeak region ($[235,295]$) even when its proportion was as low as $10^{-12}$\% of the total.

In the second step, the search method used was the so called ``full search'' (using all possible combinations  of values in each interval). The smallest possible value and the highest possible value ($minCount$ and $maxCount$ in Algorithm 2) was set with respect to the total compound's height in that given interval ($10\%$ lower and $10\%$ higher). Again, the three constituents: $^{241}$Am, $^{137}$Cs and $^{22}$Na were alternately used as missing library. When $^{137}$Cs was the missing one the LS residue was $0.0238792$, and it found:  $0.186845$ for $^{241}$Am, $0.069218$ for $^{22}$Na, $0.499977$ for $^{60}$Co, and $0.24396$ for the missing one. When  $^{241}$Am was the missing library the residue was $0.00295563$, and it found: $0.0890847$ for $^{22}$Na, $0.200009$ for $^{137}$Cs, $0.499997$ for $^{60}$Co and $0.210909$ for the missing one.  When  $^{22}$Na was chosen as the missing library, the residue was $2.29354\times 10^{-05}$ and it found: $0.183553$ for $^{241}$Am, $0.190409$ for $^{137}$Cs, $0.500001$ for $^{60}$Co and $0.126037$ for the missing one.

In the third step, the envelope was built by using the range of percentages yielded in the second step to plug into the straight lines barycentric equations, but this time using the count values of the spectrum at each channel (full spectrum instead of a 3-point). The result for $^{22}$Na with 8\% to 12\% is shown in Figure \ref{Gammaspectra}.

\begin{figure}
{\center
\includegraphics[scale=0.42]{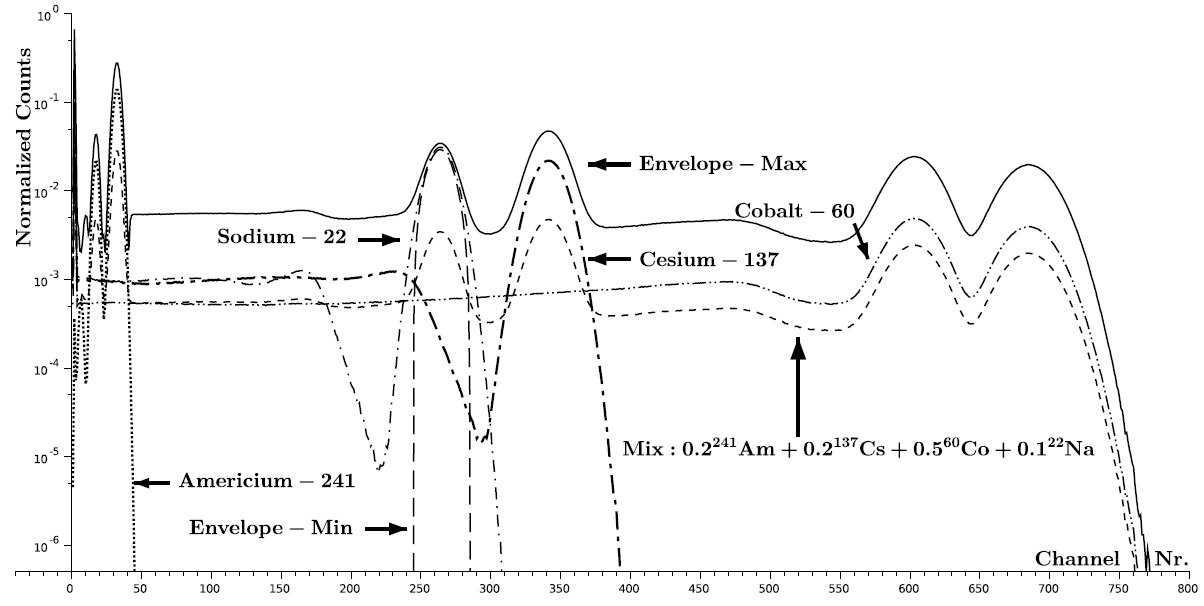}
\caption{The Monte Carlo generated gamma-ray library spectra, considering $^{22}$Na as the missing library, and the 8 to 12\%-envelope around it.}\label{Gammaspectra}
}
\end{figure}

Note that the minimum envelope approached  $^{22}$Na's spectrum only around its photopeak, while the maximum envelope approached $^{22}$Na's spectrum along a much larger interval. The geometry suggests that it is possible to reduce this envelope even further, since in order to produce the 2D region enclosed by two spectra the tetrahedron above point $P$ in Figure \ref{AffPlane} needs to be enclosed in a larger box. But the drawing of this reduced envelope is more complex due to the inequalities that define the terahedron.

\section{Conclusion}
In this work we used a geometric modeling of the spectral analysis to propose a method for the determination of a missing library among the libraries used, e.g. in the MCLLS approach. A three-step algorithm, capable of locating the missing library's photopeak, its approximate fraction and an envelope around its spectrum, was tested on Monte Carlo-generated radionuclide gamma-ray spectra. The photopeak of the missing constituent was found through the insertion of an additional coordinate whose corresponding interval of channels was searched by forcing the associated LLS system to present the highest residue. Then a new 4-vector was  added to the system to simulate the unknown, and with that an envelope was traced by using the appropriate barycentric combination. The photopeak location is found even when the missing library represents no more than $10^{-12}$\% of the total compound. The missing library's proportion was determined up to 10\% of the actual value. As for the envelope, there is room for improvement, but the fact that the missing library's photopeak was suitably enveloped represents a step forward towards identifying missing libraries in gamma-ray spectral analysis as photopeaks in a given gamma-ray spectrum are the signatures of a given radioisotope.





\section{References}
\label{}

\end{document}